# When Cyber Aggression Prediction Meets BERT on Social Media

**Zhenkun Zhou**

School of Statistics, Capital University of Economics and Business, Beijing, China

E-mail: zhenkun@cueb.edu.cn

**Mengli Yu (Corresponding author)**

School of Journalism and Communication, Nankai University, Tianjin, China

Convergence Media Research Center, Nankai University, Tianjin, China.

E-mail: mengliyu@nankai.edu.cn

**Yuxin He**

School of Statistics, Capital University of Economics and Business, Beijing, China

E-mail: heyuxin@cueb.edu.cn

**Xingyu Peng**

State Key Lab of Software Development Environment, Beihang University, Beijing, China

E-mail: xypeng@buaa.edu.cn

**Abstract:**

Increasingly, cyber aggression becomes the prevalent phenomenon that erodes the social media environment. However, due to subjective and expense, the traditional self-reporting questionnaire is hard to be employed in the current cyber area. In this study, we put forward the prediction model for cyber aggression based on the cutting-edge deep learning algorithm. Building on 320 active Weibo users' social media activities, we construct basic, dynamic, and content features. We elaborate cyber aggression on three dimensions: social exclusion, malicious humour, and guilt induction. We then build the prediction model combined with pretrained BERT model. The empirical evidence shows outperformance and supports a stronger prediction with the BERT model than traditional machine learning models without extra pretrained information. This study offers a solid theoretical model for cyber aggression prediction. Furthermore, this study contributes to cyber aggression behaviors' probing and social media platforms' organization.

**Introduction**

Aggressive behaviors traditionally are regarded as intentional act of harming others with an aroused physical state in face-to-face.[1] In the current cyber era, the explosion of information technologies and social media has changed the nature and social communication interactions. The use of social media has been a global phenomenon in human's daily life. Users will have greater spontaneity with perceived anonymity mechanism and are prone to express their feeling in the context of social networks. In this vein, social media create a new channel for users to engage in aggressive behaviors. Cyber aggression refers to the intentional harm behaviors to others through mobile phones, computers, and other electronic devices.[2,3] It includes hostile beahviors such as social exclusion, malicious humour, and guilt induction.[3,4] Past studies have explored the aggression behaviors on the popular social network platforms (e.g., Facebook and Twitter) and demonstrated that users' characteristic influence, such as age and gender.[5] For example, researchers found that males were more likely to engaging in Facebook aggression than females like sending insulting messages and posting aggressive comments.[6] Meanwhile, previous studies focused on the specific groups like adolescent and children. Pabian et al.[7] investigated the relationships between the Dark Triad personality traits and aggression among adolescents on Facebook and found that Facebook intensity significantly relates to adolescents' aggression behaviors.

It is obvious that cyber aggression will results in negative consequences, include breaking relationship, substance use, rule-breaking behaviors, and even major criminal activity.[3,8] Due to the negative influence, researchers had been trying to identify and detect cyber aggression.[8,9] As a stable individuals' psychological behavior, cyber aggression behaviors traditionally were measured by the self-reporting directly.[3] However, self-reporting weakens the data quality and validity of conclusion due to the several limitations, such as cost, subjectivity, and low flexibility.[10] The popularity of social media provides a new way to explore users' personality and psychological behaviors through amounts of data.[11] For example, Sharif and Hoque[9] proposed an aggressive text classification system to classify aggressive Bengali text into religious, gendered, verbal and political aggression classes. Sadiq et al.[8] extracted comments from social media and classified the comments text into three distinct classes of aggression. However, there is a dearth of research that strived to predict users' cyber aggression behaviors from the social media activities and personal characteristic combined with deep learning methods.

A little study had explored the cyber aggression from textual analysis.[12,13] In that, Chavan and Shylaja[12] used the traditional feature extraction techniques like TF-ID and N-gram to detect cyberbullying comments on social network. Al-Garadi et al.[13] elaborated the features' selection and tested various machine learning algorithms for prediction of cyberbullying behaviors. However, conventional machine-learning techniques were restricted in the processing natural data of their raw form that required careful engineering and domain expertise.[14] With currently development of deep neural networks, deep learning performs major advances in solving above problems. Specifically, the language model pre-training of deep learning has been shown effectively performance for processing natural language.[15] Thus, the advanced BERT (Bidirectional Encoder Representations from Transformers) model is employed in this study to predict users' cyber aggression.

Cyber aggression plays the key role in social media platforms' operations and interactions' environment. The mass data of social media and computational social science pave the advanced way to delineate individuals' personalities and behaviors. By collecting social media users' self-reports and online data, we proposed a new methodology of employing deep learning models to objectively predict individuals' cyber aggression behaviors. This study contributes detecting this phenomenon and optimizing platforms' organizations.

**Methods**

***Participants and data collection***

This study adopts the online survey methodology for data collection. Students from University of Beijing area in China are invited to participate in this survey. A total of 456 valid questionnaires are collected. All participants provide their Weibo IDs, thus we collect their Weibo activities and profiles data. To guarantee the quality of the user data, we only select active users with more than 20 Weibo tweets. Beyond that, the selected users have ever posted since 2020 and have been obviously active for at least two months in Weibo. 320 active objects are selected for this research of cyber aggression in social media, including 74 males and 246 females. We employ the Indirect Aggression Scale Aggressor version[4,5] to measure participants' cyber aggression behaviors, including three types of social exclusion (10 items), malicious humour (9 items), and guilt induction (6 items). Specifically, *social exclusion* refers to behaviors that work by socially excluding the victim, such as withholding information, leaving out of activities, and turning people against someone/ social manipulation; *malicious humour* refers to largely constituted behaviors in which humour was used to harm the victim, such as use of sarcasm as an insult, intentional embarrassment, practical joke playing; *guilt induction* refers to behaviors whereby guilt is intentionally induced, such as use of emotional blackmail, undue pressure, and coercion. Participants are introduced that "Please answer the questions below according to how you interact with your contacts on Weibo." The items are measured by 7-point Likert scales, from "1=Never behave this way" to "7=Always behave this way". Three subscales of Alpha coefficients range from 0.76 to 0.92. The social exclusion ($\mu=2.44$, $\sigma=1.11$), malicious humour ($\mu=1.69$, $\sigma=0.77$), and guilt induction ($\mu=2.03$, $\sigma=1.02$) are obtained by computing the average of relevant items ($\mu$ as Mean, $\sigma$ as SD).

**User classification**

Inspired by the work,[10] it is reasonable to divide the objects into three groups (high, neutral and low) for psychological prediction goals according to the mean and standard deviation. As an example, we label high social exclusion as category 1, neutral social exclusion as category 0 and low social exclusion as category -1. Specifically, the high social exclusion group consists of subjects with scores greater than 3.00 ($\mu + \sigma /2$), the low group consists of subjects with scores less than 1.88 ($\mu - \sigma /2$) and the neural group are those whose scores range from 1.88 to 3.00. Meanwhile, the thresholds ($\mu \pm \sigma /2$) are helpful to balance the sizes of the three categories to avoid bias in machine learning models. Based on the above method, malicious humour and guilt induction are both divided into three groups, respectively. By classifying 320 active users into three categories, we can obtain a training set to next establish machine learning models that do not require self-reports anymore.

Table 1. The breakpoints and the number of users in categories for prediction.

|  | $\mu - \sigma /2$ | $\mu + \sigma /2$ | Num of high group | Num of neural group | Num of low group |
|---|---|---|---|---|---|
| Social exclusion | 1.88 | 3.00 | 82 | 135 | 103 |
| Malicious humour | 1.30 | 2.08 | 86 | 98 | 136 |
| Guilt induction | 1.52 | 2.54 | 84 | 143 | 93 |

*Features*

As for three prediction goals social exclusion, malicious humour and guilt induction, we extract and analyze three groups of features to portray users' characteristics, and as the model inputs as well. The first and second group of features respectively contains users' basic and dynamic features. The third group charactering users' tweets content, principally depicting the text in users' tweets. Following our previous research,[11] we extract the **(1) basic features** reflect the user's demographics, static information, and elementary interactions in social media. Besides, the number and proportion of posts with pictures are added into interactive features; **(2) dynamic features** were designed to reflect the pattern of social interactions in Weibo at different time ranges. Here, social interactions included posting, mentioning, and retweeting. A vector composed of the average occurrence of the interactions at different hours or days of a week was calculated; **(3) content features** in social media can be effective indicators of psychological trait.[16,17] We adopt pretrained 300-dimention word embeddings to represent each word in Weibo, which is existed in the Pre-trained Chinese Word Vectors by Positive Pointwise Mutual Information (PPMI).[18] We average the word embeddings as the Weibo-level content vector, and then average the Weibo-level vector as the User-level 300-dimention content vectors. Emotion can be calculated from users' post tweets.[19,20] Using MoodLens model for emotion classification,[21] we classify short text into one of five emotions, including anger, disgust, happiness, sadness, and fear. For each active user, we calculate the five features that were defined as the percentages of corresponding emotional texts. Note that the text in users' description is considered into the content features. In this work, we also extract features from text content using a pretrained BERT model, which is introduced the section of Model. The extraction resulted in a 134-dimensional Basic + Dynamic feature vector and 300-dimensinal Content (word embeddings) feature vector.

*Model*

We develop three traditional machine learning models to predict the users' cyber aggression. including Logistic Regression (LR, C=1), Support vector machine (SVM, rbf kernel, C=1) and Neural Network (NN, 3 hidden layers with 128, 64 and 32 cells, 90 thousand parameters). In addition, advances in natural language processing such as BERT (Bidirectional Encoder Representation from Transformers) could enable the development of systems to potentially automate textual coding, which in turn could substantially improve psychological research. BERT is applied to generate the text represent for each social media user. it is designed to pretrain deep bidirectional representations from unlabeled text by jointly conditioning on both left and right context in all layers. This simple and powerful obtains new state-of-the-art results on eleven natural language processing tasks. Therefore, we can employ NN with pretrained BERT to predict the users' psychological characteristic, especially cyber aggression in this work. It is note that BERT is short name for the Neural Network with pretrained BERT in our study. In this work, we try to extract textual content features (embeddings) from Weibo content using a state-of-the-art modification of the BERT architecture ERNIE 3.0. This model combines auto-regressive network and auto-encoding network, so that pretrained model can handle natural language understanding through few-shot learning.[22] We encode the tweets' content via this pretrained model. In this regard, we compute the average word embeddings from the BERT as the 512-dimensinal content embeddings. In the output layer, the 134-dimensional Basic + Dynamic features are also added for training the fine-tuning to obtain the ultimate BERT model. The anonymous dataset and the code are released in the GitHub (see the section Data and code availability).

**Results**

The prediction performance of aforesaid models is verified by Accuracy (ACC), F1 score (F1) and AUC. By calculating the precision (*Pr*) and recall (*Re*) for each label *i*: $F1 = \frac{1}{3}\sum_{i=-1,0,1}\frac{2(Pr_i \cdot Re_i)}{Pr_i+Re_i}$. Moreover, Area Under the Receiver Operating Characteristic Curve (AUC) of each category against the rest categories can be measured, namely one-vs-rest AUC.[22] F1 and AUC can be the proxy for model validation even though testing dataset is not unbalanced.

Table 2 depicts the prediction performance of models for three prediction targets. As for social exclusion, the BERT model with Basic + Dynamic + Content features is outstanding, achieving ACC=62.50%, F1=53.55% and AUC=67.69%. The BERT significantly outperforms NN and other machine learning models. While comparing the results between the different groups of features, that is with and without Content features, there is no significant difference. Except for the BERT model, we compare the AUC of models with different features. AUC of LR with Basic + Dynamic features 58.55% is even a little better than that of NN with Basic + Dynamic + Content features (58.36%). It demonstrates Weibo content may not be helpful to predict people's social exclusion when the simple and traditional word embeddings are adopted. As for two another targets, malicious humour and guilt induction, we obtain the consistent results: BERT are always prime. BERT respectively improves 19.69%, 12.81% and 15.00% of ACC when compared with the second-best models.

We argue that leveraging advanced deep learning pretrained model such as BERT could potentially improve cyber aggression prediction and increase the potential of automating textual coding. Certainly, the situation that the model is built by a small batch of objects cannot be neglected.

Table 2. The performance (%) for 3-category machine learning models.

| Prediction target | Features | Model | ACC | F1 | AUC |
|---|---|---|---|---|---|
| Social exclusion | Basic + Dynamic | LR | 41.25 | 32.75 | 58.55 |
| | | SVM | 42.19 | 35.95 | 55.72 |
| | | NN | 40.94 | 34.88 | 56.40 |
| | Basic + Dynamic + Content | LR | 42.50 | 40.48 | 57.42 |
| | | SVM | 40.31 | 37.90 | 54.36 |
| | | NN | 42.81 | 40.12 | 58.36 |
| | | **BERT** | **62.50** | **53.55** | **67.69** |
| Malicious humour | Basic + Dynamic | LR | 41.88 | 31.74 | 57.87 |
| | | SVM | 42.19 | 35.43 | 56.35 |
| | | NN | 37.81 | 38.90 | 53.94 |
| | Basic + Dynamic + Content | LR | 42.50 | 38.05 | 59.47 |
| | | SVM | 43.13 | 41.41 | 58.37 |
| | | NN | 43.44 | 39.97 | 56.98 |
| | | **BERT** | **56.25** | **52.02** | **62.43** |
| Guilt induction | Basic + Dynamic | LR | 40.94 | 28.02 | 49.07 |
| | | SVM | 44.38 | 32.83 | 52.93 |
| | | NN | 37.81 | 36.55 | 51.22 |
| | Basic + Dynamic + Content | LR | 41.25 | 32.14 | 53.21 |
| | | SVM | 41.56 | 34.09 | 53.55 |
| | | NN | 39.69 | 35.74 | 53.64 |
| | | **BERT** | **59.38** | **53.76** | **64.53** |

**Discussion and conclusion**

In this study, we accurately predict human cyber aggression through social media users' activities by employing the pretrained BERT model, including predicting users' three behavioral dimensions: social exclusion, malicious humour, and guilt induction. This study provides important implications for theory and practice. From a theoretical standpoint, this paper firstly contributes to the cyber aggression. Except for the traditional self-reporting, this study answers to the call for research on cyber aggression prediction through deep learning algorithm. Second, this study offers clear guideline on BERT model prediction combining with basic, dynamic, and content features. Specifically, we found that features play the different roles in prediction of Social Exclusion, Malicious Humour, and Guilt Induction. Third, this study extends the prediction models by delineating BERT outperformance than the traditional machine learning models without external pretrained information (i.e., LR, SVM, and NN). The pretrained models indeed substantially improve social media users' behavior prediction, especially for a small size of samples in the context of psychology. Also, this study extends the theoretical prediction in the context of social media. From a pragmatic standpoint, this study offers a glimpse into a technological solution form predicting cyber aggression. It breaks the traditional limitation of self-reporting questionnaires and shed light on personality research. Furthermore, the social media platforms can monitor and recognize users' cyber aggression to avoid users' harmful behaviors. It contributes to the optimization of platforms' organizations.

However, this study has certain limitations. Firstly, more and more researchers concentrate on the interpretability of deep learning, including BERT families with attention mechanism. However, we study on the BERT model building and cyber aggression prediction, instead of model interpretability for behavior prediction, which will be a promising direction. Secondly, although deep learning BERT is employed for human behavior prediction, users' textual content is the main source to encode behavior. Next, the social media users' integrated behavior, namely basic characters, dynamic behavior and tweet content, can be a unified proxy to be feed into the behavioral encoder. This behavior encoder will build a bridge between behavior prediction and deep learning.

## Authors' Contributions

Research design, study conceptualization, model building, data analysis, and manuscript writing: ZKZ. Research design, data collection, data analysis, and manuscript writing: MLY. Model building and data analysis: YXH and XYP. Approved the submitted version for publication: ZKZ, MLY, YXH and XYP.

## Data and Code Availability Statement

The anonymous dataset and the code are available in the GitHub.
https://github.com/kayzhou/When-Cyber-Aggression-Prediction-Meets-BERT-on-Social-Media.